# Preferred Basis in Coupled Electron-Nuclear Dynamics

Junhyeok Bang[1,2,*]

[1]Department of Physics, Chungbuk National University, Cheongju 28644, Republic of Korea

[2]Research Institute for Nanoscale Science and Technology, Cheongju 28644, Republic of Korea

Beyond the adiabatic regime, our understanding of quantum dynamics in coupled systems remains limited, and the choice of representation continues to obscure physical interpretation and simulation accuracy. Here we propose a natural and efficient basis for electron–nuclear dynamics by drawing on the concepts of pointer and preferred states from decoherence theory, adapted to systems where electrons and nuclei interact strongly. Within this framework, we show that 1) the independent dynamics exploited by mixed quantum–classical (MQC) methods is best understood as a manifestation of entanglement viewed in a preferred basis, rather than a consequence of decoherence, and 2) the adiabatic Born–Oppenheimer states satisfy the conditions of an approximate preferred basis. This perspective reconciles widely used approximations with a more fundamental structure of the theory and provides a systematic route to more reliable MQC strategies. In effect, we revisit MQC methods through the lens of preferred states, clarifying when they succeed and how they can be improved.

*Corresponding authors: J. Bang (jbang@cbnu.ac.kr)

# I. Introduction

In the framework of reductionism, all materials can be regarded as coupled electron-nuclear systems. In principle, a complete solution to the time-dependent Schrödinger equation for the coupled systems would provide a comprehensive understanding of all material properties. However, this is an immensely challenging task due to the complexity of the total electron-nuclear wavefunction and the high dimensionality of the underlying Hilbert space [1,2]. In many practical cases, materials are found to operate within the adiabatic regime, where the Born-Oppenheimer approximation (BOA) allows the electronic and nuclear degrees of freedom to be treated independently [3]. Further approximating the nuclei as classical particles greatly reduces the computational burden and, when combined with modern electronic structure methods such as density functional theory, it enables realistic simulations of coupled electron-nuclear dynamics. Nevertheless, many phenomena—particularly those involving photoexcitation and electronic transitions—fall into the non-adiabatic regime, where BOA breaks down, reintroducing strong electron–nuclear correlation and the attendant complexity. To retain tractability, the community has developed mixed quantum–classical (MQC) approaches [4-9]. Yet standard implementations—such as surface hopping and Ehrenfest dynamics—are prone to known deficiencies (e.g., over-coherence, representation-dependent, detailed-balance, and branching ambiguities), often limiting predictive power for excited-state processes [10,11]. Therefore, based on a deeper understanding of the coupled electron-nuclear system, identifying an efficient and physically grounded framework for treating electron-nuclear dynamics—analogous in utility to the BOA—could lead to a new paradigm in theoretical materials science.

In this work, we aim to improve our understanding of a fundamental issue in coupled electron-nuclear dynamics: the problem of basis representation. This issue has been recognized for a long time, as exemplified by the statement "... surface-hopping is not invariant to the choice of quantum representation, ..." [12]. Nevertheless, it remains poorly understood. To shed light on this problem, we revisit a well-known example depicted in Fig. 1, originally introduced in Tully's seminal paper [4] to demonstrate the failure of the single weighted average trajectory method, also known as Ehrenfest dynamics. Here, however, we reinterpret the example from a different point of view: we show how the choice of electronic basis controls the nuclear response in MQC dynamics, which, in turn, motivates a principled search for a preferred representation capable of faithfully describing non-adiabatic processes. At the initial stage, an atom with high kinetic energy approaches a metal surface; the atom's initial nuclear state is denoted as $|i\rangle_P$, while the metal surface begins in its electronic ground state, $|g\rangle_M$, as shown in Fig. 1(a) and (b). During the collision, the surface may either remain in $|g\rangle_M$ or be promoted to an excited state $|e\rangle_M$ by absorbing kinetic energy from the atom, as illustrated in Fig. 1(c). Accordingly, two branches emerge: reflection on the ground-state potential energy surface (PES) $|g\rangle_M$ (trajectory P1) and trapping on the excited-state PES $|e\rangle_M$ (trajectory P2). Because the electronic transition is inherently probabilistic, the atom–surface system emerges in an entangled state, represented as:

$$|\Psi_i\rangle_S = |g\rangle_M |i\rangle_P \rightarrow |\Psi_f\rangle_S = a_1 |g\rangle_M |P1\rangle_P + a_2 |e\rangle_M |P2\rangle_P, \qquad (1)$$

where the coefficients $a_1$ and $a_2$ are the complex probability amplitudes associated with each outcome.

Figure 1(b) illustrates the single-weighted-average trajectory, also known as Ehrenfest dynamics. In this approach, instead of following either trajectory P1 on the ground-state surface $|g\rangle_M$ or trajectory P2 on the excited-state surface $|e\rangle_M$, the atom evolves along an intermediate path P3. This path is governed by a PES generated by the superposition of electronic states, $|\phi\rangle_M = a_1 |g\rangle_M + a_2 |e\rangle_M$, as depicted in Fig. 1(c). The resulting final state $|\Psi_f^{Ave}\rangle_S$ of the total system is therefore

$$|\Psi_f^{Ave}\rangle_S = [a_1|g\rangle_M + a_2|e\rangle_M]|P3\rangle_P = |\phi\rangle_M|P3\rangle_P, \quad (2)$$

which is a separable state, or more precisely, a parametrically separable state, in contrast to the entangled state of Eq. (1). In a parametrically separable state, the electronic and nuclear parts factorize yet share an explicit dependence on a common parameter $O_i$, i.e., $|\phi:O_i\rangle_M|P3:O_i\rangle_P$. The familiar Born–Oppenheimer (BO) adiabatic state exemplifies this form of correlation. While these states are in principle entangled [13], the correlation is relatively weak in the sense that one can treat $|\phi:O_i\rangle_M$ and $|P3:O_i\rangle_P$ independently through their mutual dependence on $O_i$ (This point will be clearly discussed below). To distinguish this weak correlation from the strongly correlated, fully entangled states in Eq. (1), we refer to such forms as *parametrically separable states* throughout this work.

Returning to the discussion of Eqs. (1) and (2), it is evident that the two final states— $|\Psi_f\rangle_S$ and $|\Psi_f^{Ave}\rangle_S$— are fundamentally inequivalent. This difference is not merely a consequence of change of basis; no unitary change of basis can convert a separable state into an entangled state and vice versa. Intuitively, one may recognize that the entangled result in Eq. (1) [Fig. 1(a)] is the physically correct [4, 14]. However, why it is correct and why the other is not correct are less obvious, and this raises a subtle yet profound question. The discrepancy originates from the choice of PES on which the atom propagates: either the ground-state surface by $|g\rangle_M$ or the excited-state surface by $|e\rangle_M$ versus PES by $|\phi\rangle_M = a_1|g\rangle_M + a_2|e\rangle_M$. Because $|\phi\rangle_M$ is itself a legitimate pure state like $|g\rangle_M$ and $|e\rangle_M$, it is not obvious *why the nucleus "chooses" the PES associated with* $|g\rangle_M$ *or* $|e\rangle_M$ *rather than the PES of* $|\phi\rangle_M$.

The puzzle deepens further when we recall that an entangled state can be rewritten in infinitely many different bases. Under a unitary change of basis, the final entangled state of Eq. (1) can be re-expressed in infinitely many equivalent forms, e.g.,

$$|\Psi_f\rangle_S = a_1|g\rangle_M|P1\rangle_P + a_2|e\rangle_M|P2\rangle_P$$
$$= b_1|A_1\rangle_M|B_1\rangle_P + b_2|A_2\rangle_M|B_2\rangle_P = \cdots, \quad (3)$$

where $\{|A_i\rangle_M\}$ and $\{|B_i\rangle_P\}$ denote alternative metal-surface basis and nuclear states, and the coefficients $b_i$ are the corresponding probability amplitudes [See Appendix I for details]. This implies that one may simulate the composite dynamics in any chosen basis—but with care. For example, if one sets $|A_1\rangle_M = |\phi\rangle_M = a_1|g\rangle_M + a_2|e\rangle_M$ as the subsystem state in Eq. (2), an exact quantum simulation yields $b_2 \neq 0$; the final state is therefore entangled, not the separable form of Eq. (2) (see Appendix I). This arbitrariness in the representation complicates the notion that "*the atom selects*" a particular electronic state: no single subsystem state is *a priori* singled out by the mathematics alone. Consequently, the earlier question "*why does the nucleus follow the PES of $|g\rangle_M$ or $|e\rangle_M$ rather than $|\phi\rangle_M$?*" must be reframed: "*1) Under what conditions does the coupled system evolve into a fully entangled state versus a (parametrically) separable one?*" Furthermore, although Eq. (3) shows mathematically equivalent representations, the physical nature—especially the time evolution of each term in the expansion—can fundamentally differ across basis choices. This leads to a second, and perhaps more practically important, question: "*2) is there a preferred basis in which the coupled dynamics are most naturally or efficiently described?*" Addressing these two questions is central to identifying a physically meaningful representation for coupled electron–nuclear dynamics.

In this work we address the two questions posed above and, on that basis, articulate a preferred basis for coupled electron–nuclear dynamics. Our analysis is grounded in the decoherence framework pioneered by H. D. Zeh, W. H. Zurek, and others, especially the notion of pointer states—basis states that remain stable under environmental monitoring [15-23]. Although decoherence has been invoked in the excited-state electron–nuclear dynamics community [24-33], it is typically used to mitigate over-coherence artifacts in MQC schemes, and explicit basis-selection criteria have not been formalized; their connection to practical MQC methods therefore remains unclear. Accordingly, Section II offers a concise review of decoherence with emphasis on pointer bases. We then highlight a key property of pointer states that, to our knowledge, has been largely overlooked: independent dynamics in a pointer-basis representation. Building on this, we argue that such independence arises from the *entanglement when expressed in a pointer basis*, rather than from decoherence per se, contrary to prior interpretations [24-33]. Section III generalizes the pointer-state concept to a preferred state suitable for interacting systems. Note that while "pointer" and "preferred" are sometimes used interchangeably in decoherence theory, we distinguish them as defined below. Section IV applies this framework to coupled electron–nuclear systems, where the subsystems interact strongly via the Coulomb force, and shows that BO states serve as an *approximate* preferred basis away from avoided crossings—both rationalizing their widespread use and delineating their regime of validity. Finally, Section V discusses efficient and accurate simulation strategies for coupled electron–nuclear dynamics within MQC schemes.

## II. Pointer States in Composite Systems and Independent Dynamics

We begin by reviewing decoherence theory and its relation to pointer states, originally developed to account for the measurement process and the quantum-to-classical transition, i.e., wavefunction collapse [19-23]. Although decoherence is invoked widely in physics, chemistry, engineering, and biology, its meaning often drifts across disciplines. In this work we adopt a narrow and precise definition of decoherence—restricted to its original formulation [19-23]— as a mechanism underlying the emergence of classicality from quantum systems. Our analysis is confined to the initial stage of decoherence, namely the formation of entanglement, which is fully described by the Schrödinger equation. We do not address the subsequent stages associated with the quantum-to-classical transition, which is not described by the Schrödinger equation. Instead of treating the full decoherence process, we limit our view within *a dynamical process in which composite systems evolve into entangled states*, which still contains its quantum nature.

Before turning to the specific case, i.e., coupled electron–nuclear systems (introduced in section IV), we begin with a generic composite system. Let the total system $\mathcal{S}$ consist of a subsystem $\mathcal{A}$ and an environment $\mathcal{E}$. Throughout this paper, we denote the states of the total system, subsystem, and environment using the subscriptions $\mathcal{S}$, $\mathcal{A}$, and $\mathcal{E}$, e.g., $|\Psi\rangle_{\mathcal{S}}$, $|\phi\rangle_{\mathcal{A}}$, and $|\chi\rangle_{\mathcal{E}}$, respectively. In later sections, $\mathcal{A}$ will represent the electronic degrees of freedom, while $\mathcal{E}$ will correspond to the nuclear degrees of freedom.

In decoherence theory, a pointer state is a subsystem state $|P_n(t_0)\rangle_{\mathcal{A}}$ that retains a separable form with the environment $\mathcal{E}$ even after interaction between the two subsystems [23]. This can be expressed as:

$$\text{Initial: } |P_n(t_0)\rangle_{\mathcal{A}} |E(t_0)\rangle_{\mathcal{E}} \rightarrow \text{Final: } \hat{U}(t,t_0)[|P_n(t_0)\rangle_{\mathcal{A}} |E(t_0)\rangle_{\mathcal{E}}] = |P_n(t)\rangle_{\mathcal{A}} |E_n(t)\rangle_{\mathcal{E}}, \quad (4)$$

where $t_0$ and $t$ denotes the initial and final times, respectively, and $\hat{U}(t,t_0)$ is the time evolution operator. The subscription *n* labels the pointer state, because, in general, multiple pointer states may exist. Interaction with different initial pointer states $|P_n(t_0)\rangle_{\mathcal{A}}$ leads to different final environment states $|E_n(t)\rangle_{\mathcal{E}}$, hence the same subscript *n* is used for both components. Here, the subscript *n* on $|E_n(t)\rangle_{\mathcal{E}}$ labels the environment 'record' correlated with the subsystem state $|P_n(t)\rangle_{\mathcal{A}}$; it is not a basis index for the environment. In general, pointer states are not required to form an orthonormal or complete set. When they do satisfy these conditions, they can be referred to as a pointer basis. In what follows, we restrict attention to cases where the pointer states form a basis set (a pointer basis), since this assumption applies to the setting analyzed in Section IV. Finally, we note that pointer states may, in general, depend on the state of the environment. This is particularly relevant in coupled electron-nuclear systems. However, to focus on the essential concepts of decoherence theory in its original form, we neglect this environment dependence in the present section and return to it in section III.

The definition of the pointer basis is intimately connected to the first question posed in the Introduction: each pointer state by construction remains separable from the environment. Consider instead a general initial state that is *not* a pointer state but rather can be represented in a superposition of pointer basis $|\phi(t_0)\rangle_\mathcal{A} = \sum_n C_n |P_n(t_0)\rangle_\mathcal{A}$, then

$$\text{Initial: } |\Psi(t_0)\rangle_S = |\phi(t_0)\rangle_\mathcal{A} |E(t_0)\rangle_\mathcal{E} = \sum_n C_n |P_n(t_0)\rangle_\mathcal{A} |E(t_0)\rangle_\mathcal{E} \rightarrow$$

$$\text{Final: } |\Psi(t)\rangle_S = \hat{U}(t, t_0)[\sum_n C_n |P_n(t_0)\rangle_\mathcal{A} |E(t_0)\rangle_\mathcal{E}] = \sum_n C_n |P_n(t)\rangle_\mathcal{A} |E_n(t)\rangle_\mathcal{E}. \quad (5)$$

The final equality holds because of the linearity of the time evolution operator $\hat{U}(t, t_0)$. Crucially, the final state is now entangled, even though the system began in a separable state. This result highlights a fundamental insight: the formation of entanglement depends solely on whether the initial state is a superposition of pointer states or not. If it is, interaction with the environment will induce entanglement. Thus, the pointer basis serves as a boundary between states that preserve separability and those that evolve into entangled composites.

Next, despite the mathematical arbitrariness in the representation, we argue that the pointer basis is the most natural and efficient basis for representing the dynamics of composite systems. In that basis, each component of the entangled superposition evolves *independently*, allowing us to track the dynamics of a single branch without accounting for interference from the others. This independence is evident in Eq. (5): the term $C_n |P_n(t)\rangle_\mathcal{A} |E_n(t)\rangle_\mathcal{E}$ evolves solely from the corresponding initial component $C_n |P_n(t_0)\rangle_\mathcal{A} |E(t_0)\rangle_\mathcal{E}$, with no contribution from any other branch [See Fig. 2(a)]. This property greatly simplifies the understanding and simulation of dynamics of composite systems.

In contrast, the situation is completely different in non-pointer representations: here, the dynamics of each term depends on all the others. Let $\{|N_n(t_0)\rangle_\mathcal{A}\}$ be a non-pointer basis at time $t_0$, such that $|N_n(t_0)\rangle_\mathcal{A} = \sum_m a_{nm}(t_0)|P_m(t_0)\rangle_\mathcal{A}$, where $a_{nm}(t_0) = {}_\mathcal{A}\langle P_m(t_0)|N_n(t_0)\rangle_\mathcal{A}$. For a basis state $|N_n(t_0)\rangle_\mathcal{A}$, the initial product $|N_n(t_0)\rangle_\mathcal{A}|E(t_0)\rangle_\mathcal{E}$ evolves as

$$\widehat{U}(t,t_0)[|N_n(t_0)\rangle_\mathcal{A}|E(t_0)\rangle_\mathcal{E}] = \sum_m a_{nm}(t_0)|P_m(t)\rangle_\mathcal{A}|E_m(t)\rangle_\mathcal{E}. \tag{6}$$

If we re-express the time evolution of the state in any non-pointer basis $\{|N_n(t)\rangle_\mathcal{A}\}$ at time $t$, we obtain

$$\sum_m a_{nm}(t_0)|P_m(t)\rangle_\mathcal{A}|E_m(t)\rangle_\mathcal{E} = \sum_{m,l} a_{nm}(t_0)a_{lm}^*(t)|N_l(t)\rangle_\mathcal{A}|E_m(t)\rangle_\mathcal{E}$$

$$= \sum_l d_{l,n}(t)|N_l(t)\rangle_\mathcal{A}|E_{l,n}^N(t)\rangle_\mathcal{E}, \tag{7}$$

where $a_{lm}^*(t) = {}_\mathcal{A}\langle N_l(t)|P_m(t)\rangle_\mathcal{A}$, and the normalized environment states $|E_{l,n}^N(t)\rangle_\mathcal{E}$ are defined via $d_{l,n}(t)|E_{l,n}^N(t)\rangle_\mathcal{E} = \sum_m a_{nm}(t_0)a_{lm}^*(t)|E_m(t)\rangle_\mathcal{E}$, with $d_{l,n}(t)$ chosen to satisfy ${}_\mathcal{E}\langle E_{l,n}^N(t)|E_{l,n}^N(t)\rangle_\mathcal{E} = 1$. As shown in Eqs. (6) and (7), an initial non-pointer basis $|N_n(t_0)\rangle_\mathcal{A}$ generally evolves into many different $|N_l(t)\rangle_\mathcal{A}$ states. Thus, for the same initial state $|\Psi(t_0)\rangle_\mathcal{S}$ in Eq. (5) but represented in the $\{|N_n(t_0)\rangle_\mathcal{A}\}$ basis, i.e., $|\Psi(t_0)\rangle_\mathcal{S} = \sum_n D_n|N_n(t_0)\rangle_\mathcal{A}|E(t_0)\rangle_\mathcal{E}$, where $D_n = \sum_m C_m a_{mn}^*(t_0)$, its time evolution becomes

$$\widehat{U}(t,t_0)[\sum_n D_n|N_n(t_0)\rangle_\mathcal{A}|E(t_0)\rangle_\mathcal{E}] = \sum_{n,l} D_n d_{l,n}(t)|N_l(t)\rangle_\mathcal{A}|E_{l,n}^N(t)\rangle_\mathcal{E}. \tag{8}$$

Each final $|N_l(t)\rangle_\mathcal{A}$ term receives contributions from multiple initial $|N_n(t_0)\rangle_\mathcal{A}$ terms [See Fig. 2(b)]. Consequently, the evolution of any given term cannot be treated independently; its dynamics are coupled to all others through quantum interference. This dynamical quantum nature, i.e., interdependence of dynamics, is a major disadvantage of using a non-pointer basis in simulating dynamics of coupled systems. Furthermore, within MQC schemes the interference embodied in Eq. (8) resides on the environment degrees of freedom, which are treated classically; enforcing such quantum interference within a classical description is inherently incompatible.

The independent dynamics property of the pointer basis provides a clear theoretical foundation for a MQC method, while it is already widely used in simulations of coupled electron-nuclear systems. For the final state $|\Psi(t)\rangle_S$ in Eq. (5), the density operator $\hat{\rho}(t)$ at time $t$ is

$$\hat{\rho}(t) = |\Psi(t)\rangle_S {}_S\langle\Psi(t)| = \sum_{n,m} C_n C_m^* |P_n(t)\rangle_{\mathcal{A}} |E_n(t)\rangle_{\mathcal{E}} {}_{\mathcal{A}}\langle P_m(t)|{}_{\mathcal{E}}\langle E_m(t)|. \qquad (9)$$

Instead of evolving the initial pure quantum state $\hat{\rho}(t_0) = |\Psi(t_0)\rangle_S {}_S\langle\Psi(t_0)|$, where $|\Psi(t_0)\rangle_S$ is given by Eq. (5), we may consider an initial classical mixture in the pointer basis: $\hat{\rho}_P^M(t_0) = \sum_n |C_n|^2 |P_n(t_0)\rangle_{\mathcal{A}} |E(t_0)\rangle_{\mathcal{E}} {}_{\mathcal{A}}\langle P_n(t_0)|{}_{\mathcal{E}}\langle E(t_0)|$. This density operator $\hat{\rho}_P^M(t_0)$ of classical mixture differs from $\hat{\rho}(t_0)$ in that it contains no coherence terms between different pointer states, yet it shares the same diagonal probabilities $|C_n|^2$ in the pointer basis. Its time evolution is

$$\hat{\rho}_P^M(t) = \hat{U}(t,t_0)\hat{\rho}_P^M(t_0)\hat{U}^\dagger(t,t_0) = \sum_n |C_n|^2 |P_n(t)\rangle_{\mathcal{A}} |E_n(t)\rangle_{\mathcal{E}} {}_{\mathcal{A}}\langle P_n(t)|{}_{\mathcal{E}}\langle E_n(t)|. \qquad (10)$$

Comparing Eq. (10) with $\hat{\rho}(t)$ in Eq. (9), we see that the diagonal elements in the pointer basis remain identical for all *t*. This means that, *within the pointer basis, the observable dynamics of the pure state $\hat{\rho}(t_0)$ and the classical mixture $\hat{\rho}_P^M(t_0)$ are equivalent*. Therefore, in simulating coupled quantum systems, one can replace the full quantum evolution with an ensemble of independently evolving subsystem–environment pairs in the pointer basis— bypassing the need to compute the full coherent dynamics. This observation underlies the rationale for a central concept of surface hopping methods, in which quantum dynamics are approximated by a mixture of classical nuclear trajectories, each evolving on a potential-energy surface determined by its corresponding electronic state.

The situation changes when the initial classical mixture is prepared in a non-pointer basis: $\hat{\rho}_N^M(t_0) = \sum_n |D_n|^2 |N_n(t_0)\rangle_{\mathcal{A}} |E(t_0)\rangle_{\mathcal{E}\ \mathcal{A}} \langle N_n(t_0)|_{\mathcal{E}} \langle E(t_0)|$. This expression is analogous to $\hat{\rho}_P^M(t_0)$ but uses the basis $\{|N_n(t_0)\rangle_{\mathcal{A}}\}$ and reproduces the same diagonal probabilities $|D_n|^2$ as the pure state $|\Psi(t_0)\rangle_S = \sum_n D_n |N_n(t_0)\rangle_{\mathcal{A}} |E(t_0)\rangle_{\mathcal{E}}$. The time evolution of $\hat{\rho}_N^M(t_0)$ is

$$\hat{\rho}_N^M(t) = \sum_{n,m,l} |D_n|^2 a_{nm}(t_0) a_{nl}^*(t_0) |P_m(t)\rangle_{\mathcal{A}} |E_m(t)\rangle_{\mathcal{E}\ \mathcal{A}} \langle P_l(t)|_{\mathcal{E}} \langle E_l(t)|. \tag{11}$$

From the relation $C_k = \sum_m D_m a_{mk}(t_0)$, the coefficient of the $k$-th diagonal term in Eqs. (9) and (10) is $|C_k|^2 = C_k C_k^* = \sum_{m,l} D_m D_l^* a_{mk}(t_0) a_{lk}^*(t_0)$, which is different with the corresponding diagonal term in Eq. (11), i.e., $\sum_n |D_n|^2 |a_{nk}(t_0)|^2$. This result clearly indicates that a MQC scheme is not invariant under a choice of basis and, when preparing a classical mixture for simulating coupled quantum dynamics, one must construct it in the pointer basis $[\hat{\rho}_P^M(t_0)]$ rather than in an arbitrary basis such as $\hat{\rho}_N^M(t_0)$. In the original decoherence theory, pointer states are those that appear in the quantum-to-classical transition [19-23,34]. Hence, representing classical objects in the pointer basis is the natural selection in MQC methods.

Note that, in the above discussion of dependent dynamics in a non-pointer basis, we introduced the initial non-pointer basis $\{|N_n(t_0)\rangle_{\mathcal{A}}\}$ for explanatory purposes only. In fact, the occurrence of dependent dynamics arises solely from using a non-pointer basis $\{|N_n(t)\rangle_{\mathcal{A}}\}$ at time $t$ — not from the choice of the initial basis. A formal proof of this statement, without invoking the initial basis $\{|N_n(t_0)\rangle_{\mathcal{A}}\}$ is provided in Appendix II.

### III. Preferred State in Interacting Systems

In the previous section, we discussed the advantages of using pointer states to represent the dynamics of composite systems. However, several important questions remain unanswered:

Do pointer states always exist? If not, what are the conditions that guarantee their existence? And when they do exist, how can we identify the pointer state for a specific system?

Furthermore, our discussion of pointer states focused on a simple conceptual setting. In realistic, strongly interacting systems, additional considerations are necessary. For example, in coupled electron–nuclear systems—where the two subsystems remain in continuous strong interaction—the dynamics of one subsystem depend sensitively on the state of the other. As a result, a subsystem state $|P_n^1(t_0)\rangle_\mathcal{A}$ that qualifies as a pointer state for a specific environment state $|E^1(t_0)\rangle_\mathcal{E}$, i.e., $\widehat{U}(t,t_0)|P_n^1(t_0)\rangle_\mathcal{A}|E^1(t_0)\rangle_\mathcal{E} = |P_n^1(t)\rangle_\mathcal{A}|E_n^1(t)\rangle_\mathcal{E}$, may fail to be a pointer state for a different environment state $|E^2(t_0)\rangle_\mathcal{E}$, instead evolving into an entangled state, i.e., $\widehat{U}(t,t_0)|P_n^1(t_0)\rangle_\mathcal{A}|E^2(t_0)\rangle_\mathcal{E} = \sum_l C_l |P_n^1(t)\rangle_\mathcal{A}|E_n^2(t)\rangle_\mathcal{E}$. Historically, decoherence theory was formulated to describe the measurement process, where the system–environment interaction is strong only in a spatially or temporally localized region, after which both systems evolve themselves in weak or negligible interaction. In those cases, the pointer-basis concept discussed in Section II is typically sufficient. In contrast, in systems where subsystems remain in a persistently strong interaction—as in coupled electron–nuclear dynamics—additional refinements to the pointer basis concept are required to account for continuous mutual influence.

To extend the pointer basis concept to more general systems, we introduce a related but more generalized notion: a state that retains the essential property of the pointer basis—independent dynamics—while relaxing other constraints. We refer to such a state as *a preferred state*, which serves as the most suitable representation for describing the dynamics of a coupled system. In terms of terminology, we draw a clear distinction between pointer states and preferred states, although the two are often used interchangeably in the literature. In this work, the term pointer basis (as discussed in Section II) retains its standard meaning within decoherence theory: the basis in which system–environment entanglement naturally suppresses coherence. The term preferred state refers to the basis that best captures the dynamics of a coupled system—particularly in scenarios involving strong and persistent interactions between subsystems.

Let us consider a specific Hermitian operator $\hat{O}$ acting on the Hilbert space of the environment. Its eigenstates $|O_i\rangle_{\mathcal{E}}$, defined by $\hat{O}|O_i\rangle_{\mathcal{E}} = O_i|O_i\rangle_{\mathcal{E}}$, form a complete and orthogonal basis for the environment's Hilbert space. We define a subsystem $|P_n:O_i\rangle_{\mathcal{A}}$ as a preferred state for the environment state $|O_i\rangle_{\mathcal{E}}$ if, for an initial total-system state $|P_n:O_i\rangle_{\mathcal{A}}|O_i\rangle_{\mathcal{E}}$ at $t_0$, the time evolution satisfies

$$|\Psi(t)\rangle_S = \hat{U}(t,t_0)[|P_n:O_i\rangle_{\mathcal{A}}|O_i\rangle_{\mathcal{E}}] = \sum_j a_{ji}^n(t)|P_n:O_j\rangle_{\mathcal{A}}|O_j\rangle_{\mathcal{E}}. \quad (12)$$

To incorporate the effects of subsystem–environment interaction while preserving independent dynamics, we allow the preferred state $|P_n:O_i\rangle_{\mathcal{A}}$ to depend parametrically on the environment label $O_i$. Although both the subsystem and environment states change during the time evolution in Eq. (12), the evolution of $|P_n:O_i\rangle_{\mathcal{A}}$ is constrained to remain correlated with the corresponding environment basis $|O_i\rangle_{\mathcal{E}}$, maintaining the same preferred-state index *n*: no *n*-sector mixes with other.

Using Eq. (12), one can confirm the independent dynamics of each *n*-th subsystem state. For a general initial state $|\Psi(t_0)\rangle_S = \sum_{ni} A_{ni}|P_n:O_i\rangle_{\mathcal{A}}|O_i\rangle_{\mathcal{E}}$, the time evolution factorizes into *n*-blocks:

$$|\Psi(t)\rangle_S = \hat{U}(t,t_0)|\Psi(t_0)\rangle_S = \sum_{nij} A_{ni}a_{ji}^n(t)|P_n:O_j\rangle_{\mathcal{A}}|O_j\rangle_{\mathcal{E}}. \quad (13)$$

Equation (14) shows that any final term $|P_n:O_j\rangle_{\mathcal{A}}$ originates *solely* from the corresponding initial *n*-th components, $\hat{U}(t,t_0)[\sum_i A_{ni}|P_n:O_i\rangle_{\mathcal{A}}|O_i\rangle_{\mathcal{E}}]$. As pointed out in pointer states, $|P_n:O_i\rangle_{\mathcal{A}}$ need not satisfy completeness or orthogonality within the subsystem Hilbert space.

If completeness and orthogonality hold for every $O_i$, we refer to the set $\{|P_n:O_i\rangle_\mathcal{A}\}$ as a preferred basis. In that case, the preferred basis associated with any $O_i$ can be used to represent any subsystem state.

In Section II, we identified two defining properties of pointer states: (i) they remain in a separable form with the environment state, and (ii) they exhibit independent dynamics for each composite pointer–environment pair. By contrast, preferred state $|P_n:O_i\rangle_\mathcal{A}$ do not satisfy the first property: they generally do not produce a strictly separable form, but instead yield an entangled state [13]. More precisely, the resulting state can be viewed as parametrically separable state due to its explicit dependence on the environment basis state $|O_i\rangle_\mathcal{E}$. As discussed above, this parametric dependence is precisely what allows preferred states to retain the second property—independent dynamics—which is the key feature for our purposes.

The above definition of $|P_n:O_j\rangle_\mathcal{A}$ provides a sufficient, but not necessary, condition for achieving independent dynamics: other states may also lead to independent dynamics with respect to the environment. Nevertheless, a key advantage of $|P_n:O_j\rangle_\mathcal{A}$ is that it offers a partial answer to the questions raised at the beginning of this section—existence and identification of preferred states. Let the total Hamiltonian of the system be $\widehat{H}^{tot} = \widehat{H}^\mathcal{A} + \widehat{H}^{int} + \widehat{H}^\mathcal{E}$, where $\widehat{H}^\mathcal{A}$ is the subsystem Hamiltonian, $\widehat{H}^\mathcal{E}$ is the environment Hamiltonian, and $\widehat{H}^{int}$ describes the interaction between them. If $\widehat{O}$ commutes with $\widehat{H}^{int}$, $[\widehat{O}, \widehat{H}^{int}] = 0$, then $\widehat{H}^{int}$ is diagonal in the environment basis $\{|O_i\rangle_\mathcal{E}\}$: $\widehat{H}^{int} = \sum_i \widehat{H}^{int}(O_i)|O_i\rangle_\mathcal{E}\,_\mathcal{E}\langle O_i|$, where $\widehat{H}^{int}(O_i)$ is an operator acting on the subsystem Hilbert space. As such, for each fixed $O_i$, we can find a parameter-dependent eigenstate $|\phi_n:O_i\rangle_\mathcal{A}$ of the subsystem: $[\widehat{H}^\mathcal{A} + \widehat{H}^{int}(O_i)]|\phi_n:O_i\rangle_\mathcal{A} = \varepsilon_n(O_i)|\phi_n:O_i\rangle_\mathcal{A}$. In the adiabatic limit, where the environment evolves slowly, the subsystem state adiabatically follows $|\phi_n:O_i\rangle_\mathcal{A}$. Its time evolution is then $\widehat{U}(t,t_0)[|\phi_n:O_i\rangle_\mathcal{A}|O_i\rangle_\mathcal{E}] =$ $\sum_j c_{ji}^n(t)|\phi_n: O_j\rangle_{\mathcal{A}}|O_j\rangle_{\mathcal{E}}$. Thus, in the adiabatic limit, $|\phi_n: O_i\rangle_{\mathcal{A}}$ can serve as the preferred state $|P_n: O_i\rangle_{\mathcal{A}}$ defined above.

For the general initial state $|\Psi(t_0)\rangle_S = \sum_{ni} A_{ni}|P_n: O_i\rangle_{\mathcal{A}}|O_i\rangle_{\mathcal{E}}$, the corresponding density operator is $\hat{\rho}(t_0) = |\Psi(t_0)\rangle_{S\,S}\langle\Psi(t_0)| = \sum_{n,m,i,k} A_{ni} A_{mk}^* |P_n:O_i\rangle_{\mathcal{A}}|O_i\rangle_{\mathcal{E}\,\mathcal{A}}\langle P_m:O_k|_{\mathcal{E}}\langle O_k|$. Its time evolution is

$$\hat{\rho}(t) = \sum_{n,m,i,j,k,l} A_{ni} A_{mk}^* a_{ji}^n(t) a_{lk}^{m*}(t) |P_n:O_j\rangle_{\mathcal{A}}|O_j\rangle_{\mathcal{E}\,\mathcal{A}}\langle P_m:O_l|_{\mathcal{E}}\langle O_l|. \tag{14}$$

The diagonal term ($n=m$ and $j=l$) is $\sum_{i,k} A_{ni} A_{nk}^* a_{ji}^n(t) a_{jk}^{n*}(t) |P_n:O_j\rangle_{\mathcal{A}}|O_j\rangle_{\mathcal{E}\,\mathcal{A}}\langle P_n:O_j|_{\mathcal{E}}\langle O_j|$. To reproduce this same diagonal term, the initial density operator for a mixture should be

$$\hat{\rho}^M(t_0) = \sum_{n,i,k} A_{ni} A_{nk}^* |P_n:O_i\rangle_{\mathcal{A}}|O_i\rangle_{\mathcal{E}\,\mathcal{A}}\langle P_n:O_k|_{\mathcal{E}}\langle O_k|, \tag{15}$$

rather than $\sum_{n,i} A_{ni} A_{ni}^* |P_n:O_i\rangle_{\mathcal{A}}|O_i\rangle_{\mathcal{E}\,\mathcal{A}}\langle P_n:O_i|_{\mathcal{E}}\langle O_i|$. This mixture $\hat{\rho}^M(t_0)$ corresponds to a classical mixture in the subsystem states $|P_n:O_i\rangle_{\mathcal{A}}$, but still contains quantum superpositions in the environment states. This observation highlights a limitation—and a caution—for MQC approaches. A detailed discussion of these implications will be provided in Section VI.

## IV. Born-Oppenheimer States: Approximate Preferred State in Coupled Electron-Nuclear Dynamics

In the preceding sections, we have highlighted the advantages of using a preferred basis to represent the dynamics of coupled systems. In this section, we focus specifically on coupled electron–nuclear systems and examine their preferred states. Our main conclusion is that the Born–Oppenheimer (BO) states can serve as an *approximate* preferred basis for such systems.

The total Hamiltonian of an electron–nuclear system $\hat{H}^T(\hat{\underline{r}}, \hat{\underline{R}})$ can be written as

$$\widehat{H}^T(\hat{\underline{r}},\hat{\underline{R}}) = \widehat{T}^N + \widehat{T}^E + \widehat{V}^E(\hat{\underline{r}}) + \widehat{V}^N(\hat{\underline{R}}) + \widehat{V}^I(\hat{\underline{r}},\hat{\underline{R}}) = \widehat{T}^N + \widehat{H}^{BO}(\hat{\underline{r}},\hat{\underline{R}}), \qquad (16)$$

where $\hat{\underline{r}}$ and $\hat{\underline{R}}$ are the sets of position operators of all electrons and nuclei, respectively. Here, $\widehat{T}^E$ and $\widehat{T}^N$ are the electronic and nuclear kinetic-energy operators; $\widehat{V}^E(\hat{\underline{r}})$ and $\widehat{V}^N(\hat{\underline{R}})$ are the electron–electron and nucleus–nucleus potential operators; and $\widehat{V}^I(\hat{\underline{r}},\hat{\underline{R}})$ is the electron–nuclear interaction potential, which plays the role of $\widehat{H}^{int}$. Notably, $[\hat{\underline{R}}, \widehat{V}^I(\hat{\underline{r}},\hat{\underline{R}})] = 0$. Following the argument in Section III, an electronic state parameterized by $\hat{\underline{R}}$ can act as a preferred state in the adiabatic limit. The operator $\widehat{H}^{BO}(\hat{\underline{r}},\hat{\underline{R}})$ is the well-known Born–Oppenheimer (BO) Hamiltonian. The energy eigenstate $|\phi_n^{BO};\underline{R}\rangle$ of $\widehat{H}^{BO}(\hat{\underline{r}};\underline{R})$ at a fixed nuclear configuration $\underline{R}$ satisfies

$$\widehat{H}^{BO}(\hat{\underline{r}};\underline{R})|\phi_n^{BO};\underline{R}\rangle = \mathcal{E}_n^{BO}(\underline{R})|\phi_n^{BO};\underline{R}\rangle, \qquad (17)$$

where $\mathcal{E}_n^{BO}(\underline{R})$ is the BO energy eigenvalue, depending parametrically on $\underline{R}$. For a given $\underline{R}$, the BO states $\{|\phi_n^{BO};\underline{R}\rangle\}$ form a complete orthonormal basis of the electronic Hilbert space. Thus, any total electron–nuclear wavefunction $|\Psi(t)\rangle$ can be expanded as $\langle\underline{R}|\Psi(t)\rangle = |\Psi(\underline{R},t)\rangle = \sum_n \chi_n(\underline{R},t)|\phi_n^{BO};\underline{R}\rangle$, where the coefficients $\chi_n(\underline{R},t)$ carry all information about the total state, encoding both nuclear and electronic system information.

To separate the contributions of the two subsystems, we decompose $\chi_n(\underline{R},t) = C_n(t)\bar{\chi}_n(\underline{R},t)$, subject to the normalization condition $\int |\bar{\chi}_n(\underline{R},t)|^2 d\underline{R} = 1$. As such, $\bar{\chi}_n(\underline{R},t)$ represents the conditional nuclear wavefunction on the electronic subsystem being in the $n$-th BO state, and $|C_n(t)|^2 = \int |\langle\phi_n^{BO};\underline{R}|\Psi(\underline{R},t)\rangle|^2 d\underline{R}$ is a purely electronic quantity, giving the probability of finding the electronic subsystem in the $n$-th BO state. This decomposition closely parallels the exact factorization approach [35,36], in which the total wavefunction is split into a conditional electronic wavefunction (subject to a normalization condition) and the nuclear wavefunction. Using $C_n(t)$ for the electronic amplitudes and $\bar{\chi}_n(\underline{R},t)$ for the normalized nuclear wavefunctions, the total state can be expressed as

$$\langle \underline{R}|\Psi(t)\rangle = |\Psi(\underline{R},t)\rangle = \sum_n C_n(t)\bar{\chi}_n(\underline{R},t)|\phi_n^{BO};\underline{R}\rangle. \tag{18}$$

Thus the equations of motion for the $C_n(t)$ and $\bar{\chi}_n(\underline{R},t)$ fully describe the dynamics of a quantum state $|\Psi(\underline{R},t)\rangle$, because $|\phi_n^{BO};\underline{R}\rangle$ is the predetermined electronic basis obtained from the BO Hamiltonian $\hat{H}^{BO}(\hat{r},\hat{\underline{R}})$.

The dynamics of the quantum state $|\Psi(\underline{R},t)\rangle$ is governed by the time-dependent Schrödinger equation: $i\hbar \frac{\partial}{\partial t}|\Psi(\underline{R},t)\rangle = \langle \underline{R}|\hat{H}^T(\hat{r},\hat{\underline{R}})|\Psi(t)\rangle$. Expanding $|\Psi(\underline{R},t)\rangle$ in the BO basis and using the decomposition above, we obtain

$$i\hbar \sum_n \left[\frac{\partial C_n(t)}{\partial t}\bar{\chi}_n(\underline{R},t) + C_n(t)\frac{\partial \bar{\chi}_n(\underline{R},t)}{\partial t}\right]\cdot|\phi_n^{BO};\underline{R}\rangle$$

$$= \sum_n \left[\left\{\sum_\alpha \frac{-\hbar^2}{2M_\alpha}C_n(t)\vec{\nabla}_\alpha^2\bar{\chi}_n(\underline{R},t) + \mathcal{E}_n^{BO}(\underline{R})C_n(t)\bar{\chi}_n(\underline{R},t)\right\}\cdot|\phi_n^{BO};\underline{R}\rangle + \right.$$

$$\left. \sum_\alpha \frac{-\hbar^2}{2M_\alpha}\left\{C_n(t)\bar{\chi}_n(\underline{R},t)\vec{\nabla}_\alpha^2|\phi_n^{BO};\underline{R}\rangle + 2C_n(t)\vec{\nabla}_\alpha\bar{\chi}_n(\underline{R},t)\cdot\vec{\nabla}_\alpha|\phi_n^{BO};\underline{R}\rangle\right\}\right], \tag{19}$$

where $\alpha$ indexes nuclei, and $M_\alpha$ is the mass of the $\alpha$-th nucleus. Projecting Eq. (19) onto $\langle \phi_n^{BO};\underline{R}|$ yields:

$$i\hbar \left[\frac{\partial C_n(t)}{\partial t}\bar{\chi}_n(\underline{R},t) + C_n(t)\frac{\partial \bar{\chi}_n(\underline{R},t)}{\partial t}\right] =$$

$$\sum_\alpha \frac{-\hbar^2}{2M_\alpha}C_n(t)\vec{\nabla}_\alpha^2\bar{\chi}_n(\underline{R},t) + \mathcal{E}_n^{BO}(\underline{R})C_n(t)\bar{\chi}_n(\underline{R},t)$$

$$+\sum_{m,\alpha}\left[\frac{-\hbar^2}{2M_\alpha}\left\{C_m(t)\bar{\chi}_m(\underline{R},t)d_{\alpha,nm}^{(2)}(\underline{R}) + 2C_m(t)\vec{\nabla}_\alpha\bar{\chi}_m(\underline{R},t)\cdot\vec{d}_{\alpha,nm}^{(1)}(\underline{R})\right\}\right] \tag{20}$$

where the first- and second-order nonadiabatic coupling terms are defined as $\vec{d}_{\alpha,nm}^{(1)}(\underline{R}) = \langle \phi_n^{BO};\underline{R}|\vec{\nabla}_\alpha|\phi_m^{BO};\underline{R}\rangle$ and $d_{\alpha,nm}^{(2)}(\underline{R}) = \langle \phi_n^{BO};\underline{R}|\vec{\nabla}_\alpha^2|\phi_m^{BO};\underline{R}\rangle$, respectively, which are functions

of $\underline{R}$. From Eq. (20), multiplying by $\bar{\chi}_n^*(\underline{R},t)$ and integrating over $\underline{R}$ gives the equation of motion for $C_n(t)$:

$$i\hbar \frac{\partial C_n(t)}{\partial t} + C_n(t) \cdot \int \bar{\chi}_n^*(\underline{R},t) \left[ i\hbar \frac{\partial}{\partial t} - \sum_\alpha \frac{-\hbar^2 \vec{\nabla}_\alpha^2}{2M_\alpha} - \mathcal{E}_n^{BO}(\underline{R}) \right] \bar{\chi}_n(\underline{R},t) \, d\underline{R} =$$

$$\sum_{m,\alpha} C_m(t) \frac{-\hbar^2}{2M_\alpha} \int \bar{\chi}_n^*(\underline{R},t) \left[ d_{\alpha,nm}^{(2)}(\underline{R}) + 2\vec{d}_{\alpha,nm}^{(1)}(\underline{R}) \cdot \vec{\nabla}_\alpha \right] \bar{\chi}_m(\underline{R},t) d\underline{R}, \quad (21)$$

and multiplying Eq. (20) by $\frac{C_n^*(t)}{|C_n(t)|^2}$ yields the equation of motion for $\bar{\chi}_n(\underline{R},t)$:

$$\left[ i\hbar \frac{\partial}{\partial t} - \sum_\alpha \frac{-\hbar^2}{2M_\alpha} \vec{\nabla}_\alpha^2 - \mathcal{E}_n^{BO}(\underline{R}) + \frac{i\hbar}{|C_n(t)|^2} C_n^*(t) \frac{\partial}{\partial t} C_n(t) \right] \bar{\chi}_n(\underline{R},t) =$$

$$\sum_{m,\alpha} \left[ \frac{-\hbar^2}{2M_\alpha} \frac{C_n^*(t) C_m(t)}{|C_n(t)|^2} \left\{ d_{\alpha,nm}^{(2)}(\underline{R}) + 2\vec{d}_{\alpha,nm}^{(1)}(\underline{R}) \cdot \vec{\nabla}_\alpha \right\} \bar{\chi}_m(\underline{R},t) \right]. \quad (22)$$

Equations (21) and (22) constitute the central coupled evolution equations for the electronic amplitudes $C_n(t)$ and the conditional nuclear wavefunctions $\bar{\chi}_n(\underline{R},t)$.

In Eq. (21), the integrand of the second term on the left-hand side evaluates to a constant. Since $\hat{H}_n^{BO}(\underline{R}) = \sum_\alpha \frac{-\hbar^2 \vec{\nabla}_\alpha^2}{2M_\alpha} + \mathcal{E}_n^{BO}(\underline{R})$ is Hermitian, and $\int \bar{\chi}_n^*(\underline{R},t) \frac{\partial}{\partial t} \bar{\chi}_n(\underline{R},t) \, d\underline{R}$ is purely imaginary due to the normalization condition $\int |\bar{\chi}_n(\underline{R},t)|^2 dR = 1$, the integrand is real and can be defined as $-E_n(t)$. The right-hand side of Eq. (21) contains coupling terms with other BO states, which drive transitions between them. In Eq. (22), the left-hand side represents a Schrödinger-like equation for $\bar{\chi}_n(\underline{R},t)$ evolving on the $n$-th single adiabatic potential $\mathcal{E}_n^{BO}(\underline{R})$, with an additional constant energy shift $-\frac{i\hbar}{|C_n(t)|^2} C_n^*(t) \frac{\partial}{\partial t} C_n(t)$. This shift is generally complex but becomes purely real when $|C_n(t)|^2$ is constant on time. The right-hand side of Eq. (22) contains the coupling terms between $\bar{\chi}_n(\underline{R},t)$ and $\bar{\chi}_m(\underline{R},t)$ on other BO states.

In many molecular systems, the non-adiabatic coupling terms $\vec{d}_{\alpha,nm}^{(1)}(\underline{R})$ and $d_{\alpha,nm}^{(2)}(\underline{R})$ vanish over most of configuration space and they become sizeable amount only in localized regions, such as near avoided crossings or conical intersections. When $\bar{\chi}_n(\underline{R},t)$ is localized in a region where both coupling terms are zero, the right-hand side of Eqs. (21) and (22) vanishes. In this case, the equations reduce to

$$i\hbar \frac{\partial C_n(t)}{\partial t} - E_n(t)C_n(t) \cong 0 \qquad (23)$$

$$\left[i\hbar \frac{\partial}{\partial t} - \sum_\alpha \frac{-\hbar^2}{2M_\alpha}\vec{\nabla}_\alpha^2 - \mathcal{E}_n^{BO}(\underline{R}) + E_n(t)\right]\bar{\chi}_n(\underline{R},t) \cong 0. \qquad (24)$$

The fourth term $E_n(t)$ in Eq. (24) is obtained from the solution of Eq. (23), $C_n(t) = C_n(t_0)\exp\left[-\frac{i}{\hbar}\int E_n(t)dt\right]$. In chemically reactive regions, avoided crossings or conical intersections may occur, coupling $C_n(t)$ and $\bar{\chi}_n(\underline{R},t)$ with those of other BO states. Once the reaction products separate and enter a region where the couplings vanish, $C_n(t)$ and $\bar{\chi}_n(\underline{R},t)$ evolve independently according to Eqs. (23) and (24). The time evolution in such regions is therefore

Initial: $|\Psi(t_0)\rangle = \bar{\chi}_n(\underline{R},t_0)|\phi_n^{BO};\underline{R}\rangle \to$

Final: $|\Psi(t)\rangle = \exp\left[-\frac{i}{\hbar}\int E_n(t)dt\right]\bar{\chi}_n(\underline{R},t)|\phi_n^{BO};\underline{R}\rangle,$ (25)

where $\bar{\chi}_n(\underline{R},t)$ is obtained solely from Eq. (24). Therefore, the BO state $|\phi_n^{BO};\underline{R}\rangle$ serves as an approximate preferred basis.

### V. Discussion

**Mass difference:** In the discussion of Sections II and III, the choice of subsystem and environment is, in principle, arbitrary. One might therefore choose the nuclear degrees of freedom as the subsystem and the electronic degrees of freedom as the environment. However, this choice is generally unfavorable because the large nuclear masses imply much smaller nuclear level spacings than electronic ones (the kinetic-energy operator scales as 1/$M$). When Eqs. (21) and (22) are formulated with the nuclei as the subsystem, the nonadiabatic coupling terms with respect to the electronic coordinates, $\vec{d}^{(1)}_{\alpha,nm}(r)$ and $d^{(2)}_{\alpha,nm}(r)$, remain nonzero over nearly the entire configuration space. The high electronic kinetic energy then mixes a broad manifold of nuclear adiabatic states, making it difficult to identify a clear preferred state for the nuclear subsystem. These considerations are fully consistent with the rationale of the Born–Oppenheimer approximation.

**Mixed Quantum-Classical method:** Based on the above discussion, we can outline the overall coupled electron–nuclear dynamics. As shown in Fig. 3, consider an initial parametrically separable state as in Eq. (25), $|\Psi(\underline{R}, t_0)\rangle = \overline{\chi}_2(\underline{R}, t_0)|\phi_2^{BO}; \underline{R}\rangle$, at $t_0$ prepared in a region where the nonadiabatic couplings vanish. In this case, the state evolves entirely within a single BO state according to Eqs. (23) and (24). Upon reaching an avoided-crossing region at time $t_1$, the state becomes coupled to other BO states, producing a fully entangled state during the interval $t_1 < t < t_2$:

$$|\Psi(\underline{R}, t_1)\rangle = C_2(t_1)\overline{\chi}_2(\underline{R}, t_1)|\phi_2^{BO}; \underline{R}\rangle \rightarrow$$

$$|\Psi(\underline{R}, t_2)\rangle = \sum_{n=1,2} C_n(t_2)\overline{\chi}_n(\underline{R}, t_2)|\phi_n^{BO}; \underline{R}\rangle. \tag{26}$$

After leaving the avoided-crossing region, each term propagates independently on its corresponding BO surface until the nuclear wavepacket $\overline{\chi}_n(R, t_3)$ reaches the next avoided-crossing region at $t_3$:

$$|\Psi(\underline{R}, t_2)\rangle = \sum_{n=1,2} C_n(t_2)\overline{\chi}_n(\underline{R}, t_2)|\phi_n^{BO}; \underline{R}\rangle \rightarrow$$

$$|\Psi(\underline{R}, t_3)\rangle = \sum_{n=1,2} C_n(t_3)\overline{\chi}_n(\underline{R}, t_3)|\phi_n^{BO}; \underline{R}\rangle, \tag{27}$$

where $|C_n(t_2)| = |C_n(t_3)|$ due to the independent propagation. Upon entering the next avoided-crossing region, *each component* of $|\Psi(\underline{R}, t_3)\rangle$ undergoes renewed nonadiabatic coupling, generating a fully entangled state as in Eq. (26). If multiple components arrive concurrently, coherent interference between them can occur (Fig. 3), thereby modifying subsequent transition probabilities and phases.

Several considerations enable efficient simulation of the processes above. In regions where the nonadiabatic couplings vanish— specifically the intervals $t_0 \to t_1$ and $t_2 \to t_3$ in Fig. 3—an effective strategy is to work in the BO (preferred) representation and propagate the conditional nuclear wavefunctions $\bar{\chi}_n(R,t)$ independently on their respective BO PESs using Eqs. (24), rather than evolving the full total wavefunction. In the absence of the branch–to–branch interference, it is often accurate to approximate the nuclear motion by classical trajectories, whose positions and momenta track $\bar{\chi}_n(R,t)$. Practically, this corresponds to propagating ensembles on each surface with weights $|C_n|^2$—the classical mixture $\hat{\rho}_P^M(t_0)$ of Eq. (10) [or $\hat{\rho}^M(t_0)$ of Eq. (15)].

In mapping quantum dynamics onto classical trajectories, several factors—often overlooked in traditional MQC methods—must be carefully considered. First, when two or more $\bar{\chi}_n(R,t)$ components overlap in the *same* avoided-crossing window, their relative phases modulate transitions among BO surfaces via quantum interference. For example (Fig. 3), if wavepackets on the upper and lower surfaces reach the crossing simultaneously at $t_3$, phase differences can alter the ensuing electronic transition. Importantly, this re-interference distinguishes the present view—independent dynamics as a property of entanglement in a preferred basis—from an interpretation based on decoherence: if independence arose from genuine decoherence (information dispersal into the environment and a quantum-to-classical transition [19-23,34]), such re-interference would not occur. Practically, one may mitigate these effects by augmenting classical propagation with auxiliary phase-evolution equations (from Eqs. (23)–(24)) that track the phases of $C_n(t)$ and the overlapping $\bar{\chi}_n(R,t)$. Second, as indicated by Eqs. (14)–(15), residual self-interference within a *single* $\bar{\chi}_n(R,t)$ can influence dynamics: two branches $\bar{\chi}_n(R_1,t_0)$ and $\bar{\chi}_n(R_2,t_0)$ may reconverge to the same configuration at time *t* and interfere. Incorporating this effect in a classical description is nontrivial; in principle, it affects ground-state as well as excited-state dynamics and warrants further study.

Second, as indicated by Eqs. (14) and (15), residual self-interference within a single $\bar{\chi}_n(R,t)$ can influence the dynamics. For example, interference arises when two nuclear wavepackets, $\bar{\chi}_n(R_1,t_0)$ and $\bar{\chi}_n(R_2,t_0)$, originating from different positions at $t_0$, reconverge to the same configuration $\bar{\chi}_n(R,t)$ at a later time *t*. This self-interference is a long-recognized hallmark of nuclear quantum effects. Incorporating it into a classical description is nontrivial; in principle, it impacts both ground- and excited-state dynamics and therefore warrants further investigation. Finally, mapping $\bar{\chi}_n(R,t)$ to classical objects is generally unfavorable in a non-preferred basis: interference between trajectories on PESs defined by a non-preferred representation must then be retained explicitly, which is incompatible with a purely classical treatment.

In the avoided-crossing regions—specifically, during $t_1 \sim t_2$ and after $t_3$ in Fig. 3—the preferred state is poorly defined, as the evolution depends sensitively on the detailed structure of the nuclear wavepacket. In such windows, a full quantum treatment of the total wavefunction may be required. Practically, these regions can be viewed as localized coupling sources that convert a *single* incoming component into *several* outgoing branches. For efficiency, one may replace explicit quantum propagation inside these windows with a reduced coupling model that delivers branching probabilities and phases. A simple option—long used in surface-hopping and Ehrenfest schemes—is to perform quantum dynamics only in the electronic subspace along prescribed classical nuclear trajectories. This suggests a hybrid strategy: employ short model "windows" at avoided crossings to estimate transition amplitudes and phases, then continue propagation of a branched ensemble on BO surfaces (e.g., via surface hopping). Such an approach combines the strengths of both descriptions while keeping computational cost manageable.

**VI. Conclusion**

We have reframed basis selection in coupled electron–nuclear dynamics through the lens of pointer and preferred states, revealing independent dynamics as a structural property of entanglement in the right representation, not as a by-product of decoherence. This insight clarifies why MQC methods work when they do, and why they fail when representation choices scramble independence. Within this framework, we showed that BO states constitute an approximate preferred basis away from avoided crossings, yielding clean, decoupled evolution of conditional nuclear wavefunctions and justifying classical nuclear propagation in those regions. Our analysis leads to a practical operating picture of nonadiabatic dynamics: *branching windows* (avoided crossings) act as localized sources of entanglement and amplitude redistribution, bracketed by extended intervals where each branch evolves independently on a single PES. This suggests hybrid algorithms that (i) estimate branching amplitudes/phases in short windows, then (ii) evolve a branched ensemble propagated classical trajectories on BO surfaces. It also explains longstanding issues in MQC—over-coherence, basis sensitivity, and branching ambiguity—as consequences of leaving the preferred representation or ignoring phase information.

ACKNOWLEDGMENT

This work was supported by Basic Science Research Program through the National Research Foundation of Korea (RS-2023-NR076774), Information Technology Research Center (IITP-RS-2024-00437284), and Global - Learning & Academic research institution for Master's·PhD students, and Postdocs (LAMP) Program of the National Research Foundation of Korea grant funded by the Ministry of Education (No. RS-2024-00445180). We used the VESTA software to generate Fig. 1 [37].

## Appendix I: Change of Basis of an Entangled State

For the final state $|\Psi_f\rangle_S = a_1|g\rangle_M|P1\rangle_P + a_2|e\rangle_M|P2\rangle_P$ of Eq. (1), one can represent it in a different but orthogonal and complete basis $|A_1\rangle_M$ and $|A_2\rangle_M$, so the $|g\rangle_M$ and $|e\rangle_M$ are represented by $|g\rangle_M = b_{g1}|A_1\rangle_M + b_{g2}|A_2\rangle_M$ and $|e\rangle_M = b_{e1}|A_1\rangle_M + b_{e2}|A_2\rangle_M$. Then

$$|\Psi_f\rangle_S = a_1[b_{g1}|A_1\rangle_M + b_{g2}|A_2\rangle_M]|P1\rangle_P + a_2[b_{e1}|A_1\rangle_M + b_{e2}|A_2\rangle_M]|P2\rangle_P$$

$$= |A_1\rangle_M[a_1 b_{g1}|P1\rangle_P + a_2 b_{e1}|P2\rangle_P] + |A_2\rangle_M[a_1 b_{g2}|P1\rangle_P + a_2 b_{e2}|P2\rangle_P]. \quad (A1)$$

Because the parentheses in the first and second terms are solely environment states, we can define $b_1|B_1\rangle_P = a_1 b_{g1}|P1\rangle_P + a_2 b_{e1}|P2\rangle_P$ and $b_2|B_2\rangle_P = a_1 b_{g2}|P1\rangle_P + a_2 b_{e2}|P2\rangle_P$, where $b_i$ is necessary to satisfy the normalization condition for $|B_i\rangle_P$. This leads Eq. (3).

## Appendix II: Re-visit dependent dynamics on a non-point basis

Equation (5) shows how a general initial state $|\Psi(t_0)\rangle_S$ evolves $|\Psi(t)\rangle_S$ independently into in the pointer basis. If we represent the final state $|\Psi(t)\rangle_S$ in any non-pointer basis $\{|N_n(t)\rangle_\mathcal{A}\}$ at time $t$, we obtain

$$|\Psi(t)\rangle_S = \sum_{n,m} C_n a^*_{mn}(t)|N_m(t)\rangle_\mathcal{A}|E_n(t)\rangle_\mathcal{E}, \quad (II.1)$$

where $a^*_{mn}(t) = {}_\mathcal{A}\langle N_m(t)|P_n(t)\rangle_\mathcal{A}$ as defined in the main text. As seen in Eq. (II.1), each $|N_m(t)\rangle_\mathcal{A}$ term originates from several initial $|P_n(t_0)\rangle_\mathcal{A}$ terms, showing the dependent dynamics in the $\{|N_m(t)\rangle_\mathcal{A}\}$ basis.

Equations (9) and (10) show that the classical mixture density operator $\hat{\rho}^M_P(t_0)$ provides equivalent dynamics of diagonal terms with the pure density operator $\hat{\rho}(t_0)$ in the pointer basis. This is not the case on the non-pointer basis. If we represent the final $\hat{\rho}(t)$ [Eq. (9)] in any non-pointer basis $\{|N_n(t)\rangle_\mathcal{A}\}$, we obtain

$$\hat{\rho}(t) = \sum_{n,l,m,k} C_n C_m^* a_{l,n}^*(t) a_{k,m}(t) |N_l(t)\rangle_\mathcal{A} |E_n(t)\rangle_\mathcal{E}\,_\mathcal{A}\langle N_k(t)|_\mathcal{E}\langle E_m(t)|. \qquad (\text{II}.2)$$

In the same way, when we represent $\hat{\rho}_P^M(t)$ in the basis,

$$\hat{\rho}_P^M(t) = \sum_{n,l,k} |C_n|^2 a_{l,n}^*(t) a_{k,n}(t) |N_l(t)\rangle_\mathcal{A} |E_n(t)\rangle_\mathcal{E}\,_\mathcal{A}\langle N_k(t)|_\mathcal{E}\langle E_n(t)|. \qquad (\text{II}.3)$$

The $l$-th diagonal terms in Eq. (II.2), i.e., $\sum_{n,m} C_n C_m^* a_{l,n}^*(t) a_{l,m}(t) |E_n(t)\rangle_\mathcal{E}\,_\mathcal{E}\langle E_m(t)|$, differ from those in Eq. (II.3), i.e., $\sum_n |C_n|^2 a_{l,n}^*(t) a_{l,n}(t) |E_n(t)\rangle_\mathcal{E}\,_\mathcal{E}\langle E_n(t)|$.

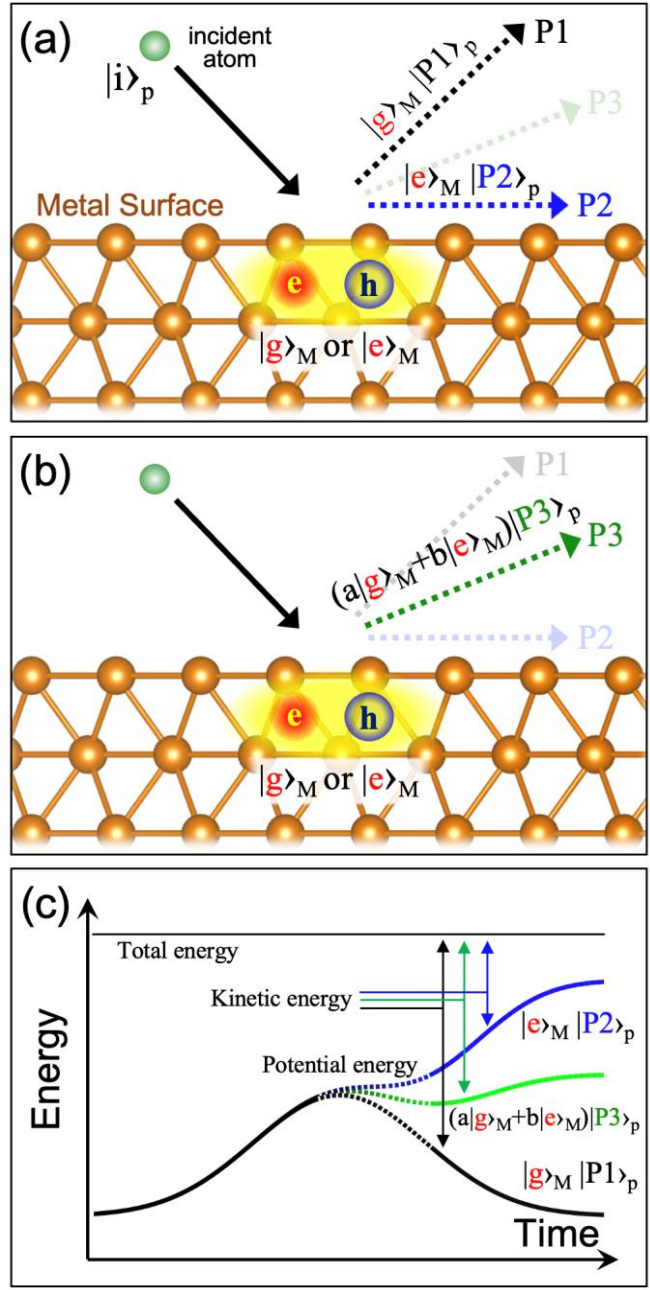

**FIG. 1.** (Color online) Schematic of an atom–metal surface scattering process showing two possible outcomes: (a) an entangled state, $|\Psi_f\rangle_S = a_1|g\rangle_M|P1\rangle_P + a_2|e\rangle_M|P2\rangle_P$, and (b) a separable state, $|\Psi_f^{Ave}\rangle_S = [a_1|g\rangle_M + a_2|e\rangle_M]|P3\rangle_P$. Initially, an atom $|i\rangle_P$ approaches the surface, and during the collision the surface remains in the ground state $|g\rangle_M$ or is promoted to the excited state $|e\rangle_M$. Accordingly, either two branches $|g\rangle_M|P1\rangle_P$ and $|e\rangle_M|P2\rangle_P$ [in (a)] or a single averaged trajectory $([a_1|g\rangle_M + a_2|e\rangle_M]|P3\rangle_P$ [in (b)] emerges. (c) Time evolution of atomic kinetic and potential energies for each trajectory: black, P1; blue, P2; green, P3.

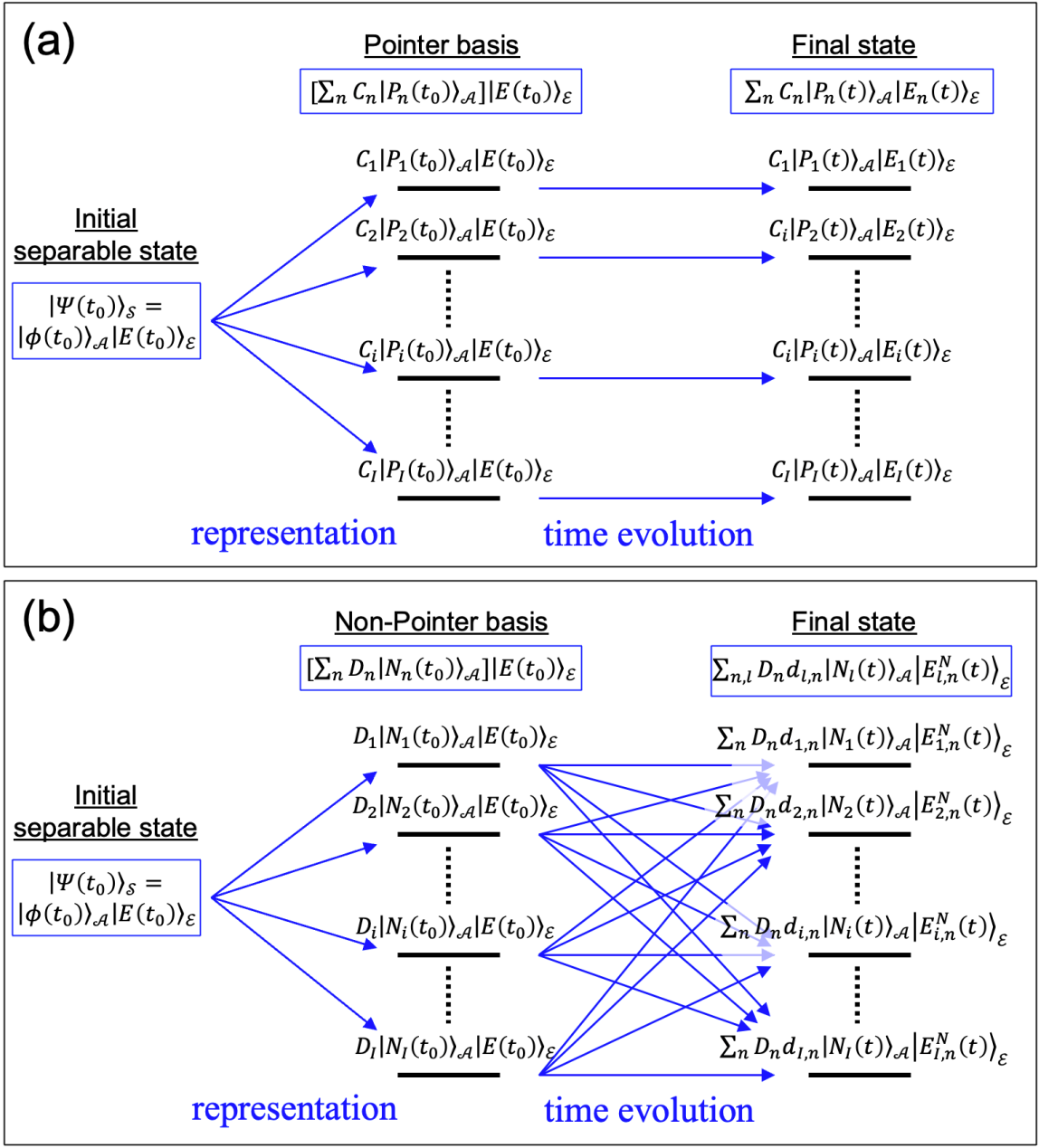

**FIG. 2.** (Color online) Schematic comparison of the time evolution of a coupled system in (a) a pointer basis and (b) a non-pointer basis. In the pointer representation, an initially separable state decomposed as $\sum_n C_n |P_n(t_0)\rangle_{\mathcal{A}} |E(t_0)\rangle_{\mathcal{E}}$ evolves to $\sum_n C_n |P_n(t)\rangle_{\mathcal{A}} |E_n(t)\rangle_{\mathcal{E}}$; each component propagates independently (no mixing between *n*). In contrast, in a non-pointer representation the same initial state $\sum_n D_n |N_n(t_0)\rangle_{\mathcal{A}} |E(t_0)\rangle_{\mathcal{E}}$ evolves to $\sum_{n,l} D_n d_{l,n}(t) |N_l(t)\rangle_{\mathcal{A}} |E^N_{l,n}(t)\rangle_{\mathcal{E}}$, exhibiting strong cross-coupling and dynamical interference among components.

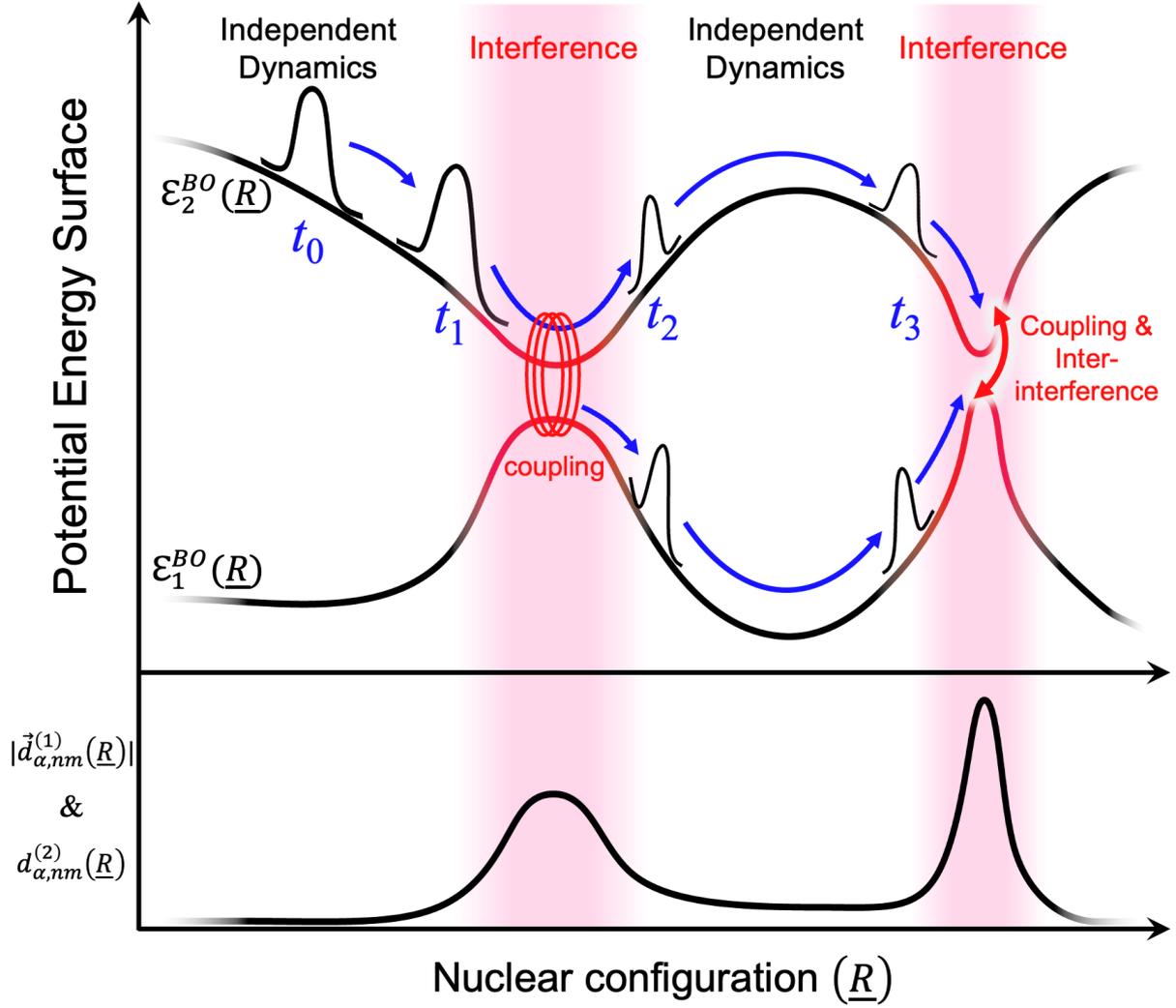

**FIG. 3.** (Color online) Schematic of coupled electron–nuclear dynamics on two BO PES $\mathcal{E}_1^{BO}(\underline{R})$ and $\mathcal{E}_2^{BO}(\underline{R})$ (top), with the magnitudes of the nonadiabatic couplings $\left|\vec{d}_{\alpha,nm}^{(1)}(\underline{R})\right|$ and $d_{\alpha,nm}^{(2)}(\underline{R})$ (bottom). At $t_0$, a parametrically separable state $|\Psi(\underline{R},t_0)\rangle = \bar{\chi}_2(\underline{R},t_0)|\phi_2^{BO};\underline{R}\rangle$ is prepared in a region where couplings vanish. When the wavepacket reaches an avoided-crossing window at $t_1$, coupling to other BO states produces a fully entangled superposition during $t_1 < t < t_2$: $|\Psi(\underline{R},t_2)\rangle = \sum_{n=1,2} C_n(t_2)\bar{\chi}_n(\underline{R},t_2)|\phi_n^{BO};\underline{R}\rangle$. Beyond the window, each component propagates independently on its respective BO surface until the next avoided-crossing region at $t_3$. Entering that region, *each branch* again undergoes coupling and leads an entangled state; if multiple branches arrive simultaneously, interference between components can occur, modifying subsequent transition probabilities and phases.